\documentclass[pdftex,twocolumn,epjc3]{svjour3}          % twocolumn
\RequirePackage[colorlinks,citecolor=blue,urlcolor=blue,linkcolor=blue]{hyperref}
\RequirePackage[T1]{fontenc}

\smartqed  % flush right qed marks, e.g. at end of proof

\RequirePackage{graphicx}
\RequirePackage{mathptmx}      % use Times fonts if available on your TeX system

\usepackage{amsmath,amssymb,amsfonts,physics,mathtools}
\usepackage{MnSymbol}
\journalname{TTMP}

\begin{document}

\title{Boson propagator under rigid rotation%\thanksref{t1}
}
\subtitle{Mode expansion approach}

\author{E. Siri\thanksref{e1}
	\and
	N. Sadooghi\thanksref{e2} %etc.
}

\thankstext{e1}{e-mail: e.siri@physics.sharif.ir}
\thankstext{e2}{e-mail: sadooghi@physics.sharif.ir}

\institute{Department of Physics, Sharif University of Technology, P.O. Box 11155-9161, Tehran, Iran \label{addr1}
}

\date{Received: date / Accepted: date}

\maketitle

\begin{abstract}
To explore how rigid rotation affects the thermodynamic properties of free relativistic bosons, we employ the standard imaginary time formalism of thermal field theory to calculate the free propagator of complex scalar fields under rotation. We introduce the corresponding partition function and explicitly compute it by expanding the modes in cylindrical coordinates. The resulting propagator is in full agreement with similar findings in the existing literature.
\end{abstract}

\section{Introduction}
\label{intro}
Recently, there has been significant interest in studying the effects of rigid rotation on the thermodynamic properties of relativistic particles \cite{Chen:2015hfc,Jiang:2016wvv,Sadooghi:2021upd,Gaspar:2023nqk,Ambrus:2023bid,Braguta:2023yjn,braguta2,Siri:2024scq,sadooghi2023}. In \cite{Ambrus:2023bid}, a system of spin zero massless bosons bounded on a one-dimensional ring is studied in the presence of an imaginary angular frequency. It is shown that the system's thermodynamics exhibits a fractal dependency on the rotation frequency, and ninionic statistics emerges. In \cite{Siri:2024scq}, an interacting relativistic bosonic gas subjected to a rigid rotation is studied at finite temperature. Various thermodynamic quantities, including pressure, energy, entropy, and angular momentum densities, as well as heat capacity, moment of inertia, and the speed of sound are computed analytically and numerically. It is shown that certain thermodynamic instabilities appear at high temperatures and large coupling constants. They are manifested as zero and negative values of the moment of inertia and heat capacity, as well as superluminal sound velocities. Zero moment of inertia, which was previously observed in \cite{Braguta:2023yjn} in a rotating and hot spin one gluon gas, leads to the phenomenon of supervorticity in relativistic a Bose gas under rigid rotation and is believed to be the reason for the negative Barnett effect \cite{braguta2}. We notice that the study of bosonic and fermionic systems under rotation may
offer potential applications in diverse fields such as heavy ion collision experiments \cite{Chen:2015hfc,Jiang:2016wvv,Sadooghi:2021upd,sadooghi2023}, condensed matter physics \cite{app-cond} and astrophysics of boson stars \cite{app-astro-1,app-astro-2,app-astro-3}. 
\par
The standard method for determining thermodynamic quantities involves calculating the partition function, which yields the pressure of relativistic gases. Other thermodynamic quantities can then be derived from the pressure. The primary component of the partition function is the free propagator. This paper aims to determine the partition function and the free propagator of relativistic bosons under rigid rotation.
\par
The propagator for free fermions under rigid rotation was first determined in \cite{Ayala:2021osy} by Ayala et al. using the generalized Fock-Schwinger method. Subsequently, in \cite{Gaspar:2023nqk}, the propagator for the Yukawa model was derived, incorporating both bosons and fermions. The general form of the boson propagator was also calculated in \cite{Siri:2024scq} using a method similar to the approaches in \cite{Ayala:2021osy} and \cite{Gaspar:2023nqk}. It is the aim of this paper, to compute the free propagator of a rotating Bose gas, by making use of an alternative method, which is referred to as `'mode expansion approach''. 
\par
The paper is organized as follows:
In Sec. \ref{sec1}, we use the corresponding metric to rigid rotation and determine the Lagrangian density of free complex scalar fields in the rigidly rotating medium.
In Sec. \ref{sec2}, we define the canonical partition function of this model and the corresponding Hamiltonian density.
Section \ref{sec3} includes the mode expansion of the boson field, including a Fourier sum in cylindrical coordinates.
In Sec. \ref{sec4}, we obtain the free propagator of rigidly rotating complex scalar fields  by using the aforementioned mode expansion.
Finally, Sec. \ref{sec5} is devoted to our conclusions.
%%%%%%%%%%%%%%%%%%%%%%%%%%%%%
\section{Free complex field in the presence of rigid rotation}
\label{sec1}
We use the Lagrangian density (\ref{Lagrangian}) to describe the non-interacting relativistic bosons,
\begin{equation}\label{Lagrangian}
	\mathcal{L}=g^{\mu\nu}\partial_{\mu}\phi^{*}\partial_{\nu}\phi - m^{2} \phi^{*}\phi,
\end{equation}
where
\begin{equation}\label{rmetric}
	{{g}_{\mu \nu }}=\left( \begin{matrix}
	1-\left( {{x}^{2}}+{{y}^{2}} \right){{\Omega }^{2}} & y\Omega  & -x\Omega  & 0  \\
	y\Omega  & -1 & 0 & 0  \\
	-x\Omega  & 0 & -1 & 0  \\
	0 & 0 & 0 & -1  \\
\end{matrix} \right),
\end{equation}
is a rigid rotation metric \cite{Chen:2015hfc}. Considering this metric means that the bosons rotate around the $z$-axis with a uniform angular velocity $\Omega$. By defining angular momentum
\begin{equation}\label{angularmom}
	L_{z}=i(y\partial_{x}\phi-x\partial_{y}\phi),
\end{equation}
and plugging (\ref{rmetric}) into (\ref{Lagrangian}), the Lagrangian density is expressed as
\begin{equation}\label{BEC-eq4}
	\mathcal{L} = (\partial_{0}-i\Omega L_{z} )\phi^{*}(\partial_{0}-i\Omega L_{z})\phi - \left| \grad{\phi} \right|^{2} - m^2 \left| \phi \right|^{2}.
\end{equation}
Also the conjugate momentum of each fields can be obtained form Lagrangian density and are given by
\begin{subequations}
	\begin{align}
		\pi&=\frac{\partial\mathcal{L}}{\partial(\partial_{0}\phi^{*})} =\partial_{0}\phi-i\Omega L_{z} \phi \label{5a}, \\
		\pi^{*}&=\frac{\partial\mathcal{L}}{\partial(\partial_{0}\phi)} =\partial_{0}\phi^{*}-i\Omega L_{z} \phi^{*}. \label{5b}
	\end{align}
\end{subequations}
%%%%%%%%%%%%%%%%%%%%%%%%%%%%
\section{Partition function}
\label{sec2}
%%%%%%%%%%%%%%%%%%%%
In the canonical ensemble, the partition function is given by
\begin{equation}\label{BEC-eq14}
	\mathcal{Z} = \int \mathcal{D}\pi^{*}\mathcal{D}\pi \int \mathcal{D}\phi^{*}\mathcal{D}\phi \exp \left[\int_{X} \mathcal{J}\left(\pi^{*},\pi^{*};\phi^{*},\phi\right) \right],
\end{equation}
where
\begin{equation}\label{BEC-eq15}
	\int_{X} \equiv \int_{0}^{\beta} d\tau \int d^{3}x, \qquad\text{and}\qquad \beta \equiv 1/T.
\end{equation}
Here, $\tau=it$ is imaginary time \cite{Laine:2016hma}. In (\ref{BEC-eq14}), the integrand is defined as
 \begin{equation}\label{BEC-eq8}
	\mathcal{J}\left(\pi^{*},\pi^{*};\phi^{*},\phi\right)\equiv(\pi^{*}\partial_{0}\phi+\pi \partial_{0}\phi^{*}-\mathcal{H}),
\end{equation}
where
\begin{equation}\label{BEC-eq9}
\mathcal{H} = \partial_{0}\phi^{*} \pi+ \partial_{0}\phi \pi^{*} - \mathcal{L},
\end{equation}
denotes the Hamiltonian density. Plugging (\ref{BEC-eq4}) into  (\ref{BEC-eq9}) and using (\ref{5a}) and (\ref{5b}), yields
\begin{equation}\label{BEC-eq10}
	\mathcal{H}=\left| \pi \right|^{2}+i\Omega(\pi L_{z} \phi^{*}+\pi^{*} L_{z} \phi) + \left| \grad{\phi} \right|^{2} + m^{2} \left| \phi \right|^{2}.
\end{equation}
To integrate over $\pi$ and $\pi^{*}$ in \eqref{BEC-eq14}, we introduce following shifted momenta:
\begin{subequations}
	\begin{eqnarray}
		\tilde{\pi} &=& \pi-\partial_{0}\phi+i\Omega L_{z}\phi, \\	
		\tilde{\pi}^{*} &=& \pi^{*}-\partial_{0}\phi^{*}+i\Omega L_{z}\phi^{*}.
	\end{eqnarray}
\end{subequations}
Using (\ref{BEC-eq10}) and after some standard algebraic steps, the integrand (\ref{BEC-eq8}) reads
\begin{equation}
		\mathcal{J} =  -\tilde{\pi}^{*}\tilde{\pi}+\mathcal{L'},
\end{equation}
where $	\mathcal{L'}$ is given by
\begin{equation}
		\mathcal{L'} = \left| (\partial_{0}-i\Omega L_{z}) \phi \right|^{2} - \left| \grad{\phi} \right|^{2} - m^{2} \left| \phi \right|^{2}.
\end{equation}
%%%%%%%%%%%%%
	\section{Mode expansion in cylindrical coordinates}
	\label{sec3}
	We can describe a complex scalar field \(\phi(x)\) with a periodic boundary condition \(\phi(\tau=0) = \phi (\tau= \beta)\) by using a Fourier sum in cylindrical coordinates $(\tau,r,\varphi,z)$,
\begin{subequations}\label{BEC-eq21}
	\begin{align}
		\phi(x)&=\sqrt{\beta V}\sumint_{n,\ell,k} e^{i(\omega_{n}\tau+\ell \varphi + k_{z}z)} J_{\ell}(k_{\perp}r) \tilde{\phi}_{n,\ell}(k) \label{14a}, \\
		\phi^{*}(x)&=\sqrt{\beta V}\sumint_{n',\ell',k'} e^{-i(\omega_{n'}\tau+\ell' \varphi + k'_{z}z)} J_{\ell'}(k'_{\perp}r) \tilde{\phi}^{*}_{n',\ell'}(k'). \label{14b}
	\end{align}
\end{subequations}
Here, $r$ is the radial
coordinate, $\varphi$ the azimuthal angle, and $z$ the height of
the cylinder. Moreover, $\omega_{n} \equiv 2 \pi n T$ with $n\in \mathbb{Z}$  are Matsubara frequencies and $J_{\ell}(k_{\perp}r)$ refers to Bessel function of the first kind. Following notation is introduced,
\begin{subequations}\label{BEC-eq23}
	\begin{align}
		\sumint_{n,\ell,k} &\equiv \sum_{n=-\infty}^{+\infty}\sum_{\ell=-\infty}^{+\infty} \int\frac{dk_{\perp}k_{\perp}dk_{z}}{(2\pi)^{2}},  \\
		\sumint_{n',\ell',k'} &\equiv \sum_{n'=-\infty}^{+\infty}\sum_{\ell'=-\infty}^{+\infty} \int\frac{dk'_{\perp}k'_{\perp}dk'_{z}}{(2\pi)^{2}}.
	\end{align}
\end{subequations}
The current series expansion differs from the conventional expansion introduced in references such as \cite{Kapusta:2006pm,Laine:2016hma} where it includes solutions derived from the Klein-Gordon equation in cylindrical coordinates. In this coordinate system, these solutions involve plane waves in three directions and Bessel functions along the radial direction \cite{Siri:2024scq}.
%%%%%%%%%%%%%%%%%%%%%%
\section{Free propagator under rigid rotation}
\label{sec4}
According to above arguments, the partition function of free complex scalar fields is given by
	\begin{equation}\label{BEC-eq24}
	\mathcal{Z} = \int \mathcal{D}\pi^{*}\mathcal{D}\pi \int \mathcal{D}\phi^{*}\mathcal{D}\phi \exp \left[\int_{X} (\mathcal{L'}-\tilde{\pi}^{*}\tilde{\pi}) \right].
\end{equation}
Performing the Gaussian integration over shifted momenta, we arrive at
\begin{equation}\label{BEC-eq25}
	\mathcal{Z} = N \int \mathcal{D}\phi^{*}\mathcal{D}\phi \exp \left[\int_{X} \mathcal{L'} \right] .
\end{equation}
The result of the Gaussian momentum integral is included in the constant factor $N$.
In cylindrical coordinates, (\ref{BEC-eq15}) turns out to be
\begin{equation}\label{BEC-eq26}
	\int_{X} = \int_{0}^{\beta} d\tau \int_{0}^{\infty} dr \, r \int_{0}^{2\pi}d\varphi \int_{-\infty}^{\infty} dz.
\end{equation}
In what follows, we compute
\begin{equation}\label{BEC-eq27}
	\int_{X} \mathcal{L'} = 	\int_{X} \left| (\partial_{0}-i\Omega L_{z}) \phi \right|^{2} - \left| \grad{\phi} \right|^{2} - m^{2} \left| \phi \right|^{2},
\end{equation}
by using the mode expansion of (\ref{14a}) and (\ref{14b}). Utilizing the completeness relations presented in Appendix \ref{app1}, it can be readily demonstrated that
\begin{equation}\label{BEC-eq20}
	\begin{aligned}
		I &= \int_{X} e^{i(\omega_{n}-\omega_{n'})\tau}  e^{i(\ell-\ell')\varphi} e^{i(k_{z}-k'_{z})z} J_{\ell}(k_{\perp}r)  J_{\ell}(k'_{\perp}r) \\
		&=\beta (2\pi)^{2} \hat{\delta}_{\ell,\ell'}^{n,n'}(k_{z},k_{\perp};k'_{z},k'_{\perp}),
	\end{aligned}
\end{equation}
where $\hat{\delta}_{\ell,\ell'}^{n,n'}$ is defined as
\begin{equation}\label{BEC-eq30}
	\hat{\delta}_{\ell,\ell'}^{n,n'}(k_{z},k_{\perp};k'_{z},k'_{\perp})\equiv \frac{1}{k_{\perp}} \delta(k_{z}-k'_{z}) \delta(k_{\perp}-k'_{\perp}) \delta_{n,n'} \delta_{\ell,\ell'}.
\end{equation}
The first contribution of \eqref{BEC-eq27} is first given by
\begin{equation}\label{BEC-eq45}
	\mathcal{I}_{1}\equiv \int_{X} \left| (\partial_{0}-i\Omega L_{z}) \phi \right|^{2} = \int_{X} (\partial_{0}-i\Omega L_{z})\phi^{*}(\partial_{0}-i\Omega L_{z})\phi,
\end{equation}
where in the imaginary time formalism $\partial_{0}= i \partial_{\tau}$. Plugging the mode expansion of $\phi$ and $\phi^{*}$ into  \eqref{BEC-eq45}, using
\begin{eqnarray}\label{BEC-eq46}
		(i \partial_{\tau}-i\Omega L_{z})\phi^{*} &=&\sqrt{\beta V} \sumint_{n',\ell',k'} \left( \omega_{n'} + i \ell' \Omega  \right) e^{-i(\omega_{n'}\tau+\ell' \varphi + k'_{z}z)}\nonumber\\
	&&\times  J_{\ell'}(k'_{\perp}r) \tilde{\phi}^{*}_{n',\ell'}(k'),
\end{eqnarray}
and
\begin{equation}\label{BEC-eq47}
	\begin{aligned}
		(i \partial_{\tau}-i\Omega L_{z})\phi &=\sqrt{\beta V} \sumint_{n,\ell ,k}  \left( -\omega_{n} + i \ell \Omega  \right) e^{i(\omega_{n}\tau+\ell \varphi + k_{z}z)} \\
		& \times J_{\ell}(k_{\perp}r) \tilde{\phi}_{n,\ell}(k),
	\end{aligned}
\end{equation}
as well as (\ref{BEC-eq20}), and performing the integration over $k'$ and summation over $n',\ell'$, we have
\begin{equation}\label{BEC-eq49}
	\mathcal{I}_{1}	= - V \sumint_{n,\ell ,k} \tilde{\phi}^{*}_{n,\ell}(k) \, \left(\beta^{2}\left(\omega_{n} + i \ell \Omega \right)^{2}\right) \, \tilde{\phi}_{n,\ell}(k).
\end{equation}
The gradient contribution of (\ref{BEC-eq27}) is
\begin{equation}\label{BEC-eq32}
	\mathcal{I}_{2}\equiv\int_{X} \left|\grad{\phi}\right|^{2} = \int_{X} \grad{\phi^{*}}\grad{\phi},
\end{equation}
where the gradient operator in the cylindrical coordinates is given by
\begin{equation}\label{BEC-eq33}
	\grad{\phi}=\frac{\partial \phi}{\partial r}\hat{r}+\frac{1}{r}\frac{\partial \phi}{\partial \varphi}\hat{\varphi}+\frac{\partial \phi}{\partial z}\hat{z}.
\end{equation}
Using the orthonormality of the unit vectors in cylinder coordinates, (\ref{BEC-eq32}) is equal to
\begin{equation}\label{BEC-eq34}
	\mathcal{I}_{2} = \int_{X} (\frac{\partial \phi^{*}}{\partial r}\frac{\partial \phi}{\partial r}+\frac{1}{r^{2}}\frac{\partial \phi^{*}}{\partial \varphi} \frac{\partial \phi}{\partial \varphi}+\frac{\partial \phi^{*}}{\partial z}\frac{\partial \phi}{\partial z}).
\end{equation}
The radial derivative is
\begin{subequations} \label{BEC-eq35}
	\begin{align}
		\frac{\partial \phi^{*}}{\partial r} &=\sqrt{\beta V}\sumint_{n',\ell',k'} e^{-i(\omega_{n'}\tau+\ell' \varphi + k'_{z}z)} \frac{\partial J_{\ell'}(k'_{\perp}r)}{\partial r}  \tilde{\phi}^{*}_{n',\ell'}(k'), \\
		\frac{\partial \phi}{\partial r} &=\sqrt{\beta V} \sumint_{n,\ell,k} e^{i(\omega_{n}\tau+\ell \varphi + k_{z}z)} \frac{\partial J_{\ell}(k_{\perp}r)}{\partial r}  \tilde{\phi}_{n,\ell}(k).
	\end{align}
\end{subequations}
Using the relations of Bessel functions from Appendix \ref{app2}, the product of radial derivatives is given by
\begin{eqnarray}\label{BEC-eq40}
\lefteqn{
		\frac{\partial \phi^{*}}{\partial r}\frac{\partial \phi}{\partial r} =\beta V \sumint_{n',\ell' ,k'} \sumint_{n,\ell,k} e^{-i(\omega_{n'}\tau+\ell' \varphi + k'_{z}z)}  e^{i(\omega_{n}\tau+\ell \varphi + k_{z}z)} }\nonumber\\
		&&\times \frac{k_{\perp}k'_{\perp}}{4} \bigg[ 2J_{\ell-1}(k_{\perp}r)  J_{\ell'-1}(k'_{\perp}r) +2  J_{\ell+1}(k_{\perp}r)  J_{\ell'+1}(k'_{\perp}r)
		 \nonumber\\
		&&   -\frac{4 \ell \ell'}{k_{\perp}k'_{\perp}r^{2}} J_{\ell}(k_{\perp}r)  J_{\ell'}(k'_{\perp}r) \bigg] \tilde{\phi}^{*}_{n',\ell'}(k') \tilde{\phi}_{n,\ell}(k).
\end{eqnarray}
The azimuthal derivative is computed as follows:
\begin{subequations} \label{BEC-eq41}
	\begin{align}
		\frac{\partial \phi^{*}}{\partial \varphi} &=\sqrt{\beta V}\sumint_{n',\ell' ,k'} \ell'\, e^{-i(\omega_{n'}\tau+\ell' \varphi + k'_{z}z)} J_{\ell'}(k'_{\perp}r) \tilde{\phi}^{*}_{n',\ell'}(k'),\\
		\frac{\partial \phi}{\partial \varphi} &=\sqrt{\beta V} \sumint_{n,\ell ,k} \ell \, e^{i(\omega_{n}\tau+\ell \varphi + k_{z}z)}  J_{\ell}(k_{\perp}r) \tilde{\phi}_{n,\ell}(k).
	\end{align}
\end{subequations}
The product of azimuthal derivative is thus given by
\begin{eqnarray}\label{BEC-eq42}
		\lefteqn{\frac{\partial \phi^{*}}{\partial \varphi}\frac{\partial \phi}{\partial \varphi} =\beta V \sumint_{n',\ell' ,k'} \sumint_{n,\ell,k} \ell \ell' \, e^{-i(\omega_{n'}\tau+\ell' \varphi + k'_{z}z)}  e^{i(\omega_{n}\tau+\ell \varphi + k_{z}z)} }\nonumber\\
		&& \times J_{\ell}(k_{\perp}r)  J_{\ell'}(k'_{\perp}r)  \tilde{\phi}^{*}_{n',\ell'}(k') \tilde{\phi}_{n,\ell}(k).	
\end{eqnarray}
Finally, the product of derivative in the $z$ direction is given by
\begin{eqnarray}\label{BEC-eq43}
		\lefteqn{\frac{\partial \phi^{*}}{\partial z}\frac{\partial \phi}{\partial z} =\beta V\sumint_{n',\ell' ,k'} \sumint_{n,\ell,k} k_{z} k'_{z} \, e^{-i(\omega_{n'}\tau+\ell' \varphi + k'_{z}z)}  e^{i(\omega_{n}\tau+\ell \varphi + k_{z}z)} }\nonumber\\
		&& \times J_{\ell}(k_{\perp}r)  J_{\ell'}(k'_{\perp}r)  \tilde{\phi}^{*}_{n',\ell'}(k') \tilde{\phi}_{n,\ell}(k).
\end{eqnarray}
Similarly, by using (\ref{BEC-eq20}) and performing integration over $k'$ and summation over $n',\ell'$, we arrive at
\begin{equation}\label{BEC-eq44}
	\mathcal{I}_{2}	=  V \sumint_{n,\ell ,k} \tilde{\phi}^{*}_{n,\ell}(k) \, \left(\beta^{2}\left(k^{2}_{\perp}+k^{2}_{z}\right)\right) \, \tilde{\phi}_{n,\ell}(k).
\end{equation}
The final result for the last contribution is given by
\begin{equation}\label{BEC-eq31}
	 \mathcal{I}_{3}= V \sumint_{n,\ell ,k} \tilde{\phi}^{*}_{n,\ell}(k) \left(\beta^{2}m^{2}\right)  \tilde{\phi}_{n,\ell}(k).
\end{equation}
Combining equations (\ref{BEC-eq49}), (\ref{BEC-eq44}), and (\ref{BEC-eq31}), we obtain
\begin{equation}\label{BEC-eq50}
	\int_{X} \mathcal{L'} = - V \sumint_{n,\ell ,k} \tilde{\phi}^{*}_{n,\ell}(k) \, \left(\beta^{2} D_{\ell,0}^{-1}(k)\right) \, \tilde{\phi}_{n,\ell}(k),
\end{equation}
where 	$D_{\ell,0}^{-1}(k) $ is the inverse free propagator in the presence of rigid rotation,
\begin{equation}\label{BEC-eq51}
	D_{\ell,0}^{-1}(k) \equiv (\omega_{n}+i\ell\Omega)^{2}+\omega^{2},
\end{equation}
and $\omega^{2}=k^{2}_{\perp}+k^{2}_{z}+m^{2}$.

\section{Conclusion}
\label{sec5}
In this paper, we have derived the expression (\ref{BEC-eq51}) representing the propagator of a free boson at finite temperature under the influence of a rigid rotation. Our approach utilized a new method based on mode expansion in the cylindrical coordinate system, deviating from the standard plane wave approach used in references such as \cite{Kapusta:2006pm} and \cite{Laine:2016hma}. The previously calculated propagator for a rigidly rotating Bose gas at finite temperature, as seen in \cite{Gaspar:2023nqk} and \cite{Siri:2024scq}, was determined using the generalized Fock-Schwinger method. The result in \eqref{BEC-eq51} is consistent with the findings in \cite{Siri:2024scq}, except for the sign of $\ell$, which is related to the symmetric nature of this propagator under such a change (specifically, see Eq. (2.25) in \cite{Siri:2024scq}).
It is worth noting that the Fock-Schwinger method emphasizes geometric and path integral aspects and uses the Schwinger proper time technique, while the mode expansion approach focuses on the Fourier transform and operator formalism. In this work, a Bessel-Fourier expansion is employed due to the special cylindrical geometry. Each of these methods has its own advantages. For example, the Fock-Schwinger method is more advantageous in curved space-time scenarios, while mode expansion is generally more straightforward for perturbative calculations in flat space-time. Importantly, both methods yield the same result, as demonstrated in the present study.
\section{Acknowledgments}
This work is presented by E. S. at the 8th Iranian Conference on Mathematical Physics (ICMP) in Qom, Iran, on 7-8 July 2024. The authors express their gratitude to the conference organizers for the invitation and hospitality.
\appendix

\section{Useful completeness relations}\label{app1}
To arrive at (\ref{BEC-eq20}), we used completeness relations \cite{Siri:2024scq}
\begin{subequations}\label{BEC-eq28}
	\begin{align}
		\int_{0}^{\beta} d\tau e^{i(\omega_{n}-\omega_{n'})\tau} &=\beta\delta_{n,n'},  \\
		\int_{0}^{2\pi} d\varphi e^{i(\ell-\ell')\varphi} &= (2\pi) \delta_{\ell,\ell'}, \label{BEC-eq28b} \\
		\int_{-\infty}^{\infty} dz e^{i(k_{z}-k'_{z})z} &= (2\pi) \delta(k_{z}-k'_{z}),  \\
		\int_{0}^{\infty} dr\, r J_{\ell}(k_{\perp}r)  J_{\ell}(k'_{\perp}r) &=\frac{1}{k_{\perp}} \delta(k_{\perp}-k'_{\perp}).
	\end{align}
\end{subequations}
\section{Bessel function relations}\label{app2}
To derive the propagator in Fourier space, we used, in particular, the derivative of the $\ell$-th order Bessel function
\begin{equation}\label{BEC-eq36}
	\frac{\partial J_{\ell}(k_{\perp}r)}{\partial r}
	= \frac{k_{\perp}}{2} \left( J_{\ell-1}(k_{\perp}r) - J_{\ell+1}(k_{\perp}r)  \right),
\end{equation}
and another useful recursive relation \cite{Abramowitz2006}
\begin{equation}\label{BEC-eq37}
	J_{\ell+1}(k_{\perp}r) = \frac{2\ell}{k_{\perp}r}J_{\ell}(k_{\perp}r)-J_{\ell-1}(k_{\perp}r).
\end{equation}


\begin{thebibliography}{99}
\bibitem{Chen:2015hfc}
H.~L.~Chen, K.~Fukushima, X.~G.~Huang and K.~Mameda,
Phys. Rev. D \textbf{93}, 10, 104052 (2016)
%%%%%%%%%%%%%%%%%%%%%%%%%%%%%%%%%%%%
\bibitem{Jiang:2016wvv}
Y.~Jiang and J.~Liao,
Phys. Rev. Lett. \textbf{117}, 19, 192302 (2016)
%%%%%%%%%%%%%%%%%%%%%%%%%%%%%%%%%%%%
\bibitem{Sadooghi:2021upd}
N.~Sadooghi, S.~M.~A.~Tabatabaee Mehr and F.~Taghinavaz,
Phys. Rev. D \textbf{104}, 11, 116022 (2021)
%%%%%%%%%%%%%%%%%%%%%%%%%%%%%%%%%%%%
\bibitem{Gaspar:2023nqk}
I.~I.~Gaspar, L.~A.~Hern\'andez and R.~Zamora,
Phys. Rev. D \textbf{108}, 9, 094020 (2023)
%%%%%%%%%%%%%%%%%%%%%%%%%%%%%%%%%%%%
\bibitem{Ambrus:2023bid}
V.~E.~Ambru\c{s} and M.~N.~Chernodub,
Phys. Rev. D \textbf{108}, 8, 085016 (2023)
%%%%%%%%%%%%%%%%%%%%%%%%%%%%%%%%%%%%
\bibitem{Braguta:2023yjn}
V.~V.~Braguta, M.~N.~Chernodub, A.~A.~Roenko and D.~A.~Sychev,
Phys. Lett. B \textbf{852}, 138604 (2024)
%%%%%%%%%%%%%%%%%%%%%%%%%%%%%%%%%%%%
\bibitem{braguta2}
V.~V.~Braguta, M.~N.~Chernodub, I.~E.~Kudrov, A.~A.~Roenko and D.~A.~Sychev,
Phys. Rev. D \textbf{110}, 014511 (2024)
%%%%%%%%%%%%%%%%%%%%%%%%%%%%%%%%%%%%
\bibitem{Siri:2024scq}
E.~Siri and N.~Sadooghi,
Phys. Rev. D \textbf{110}, 036016 (2024)
%%%%%%%%%%%%%%%%%%%%%%%%%%%%%%%%%%%%
\bibitem{sadooghi2023}
H.~Mortazavi~Ghalati and N.~Sadooghi,
Phys. Rev. D \textbf{108}, 054032 (2023)
%%%%%%%%%%%%%%%%%%%%%%%%%%%%%%%%%%%%
\bibitem{app-cond}
J. Lon$\check{c}$ar, B. Igrec and D. Babi$\acute{c}$, Symmetry \textbf{14}, 529 (2022)
%doi.org/10.3390/sym14030529
%%%%%%%%%%%%%%%%%%%%%%%%%%
\bibitem{app-astro-1}
F.~E.~Schunck and E.~W.~Mielke,
Phys. Lett. A \textbf{249}, 389 (1998)
%doi:10.1016/S0375-9601(98)00778-6
%%%%%%%%%%%%%%%%%%%%%%%%%%
\bibitem{app-astro-2}
F.~E.~Schunck and E.~W.~Mielke,
Class. Quant. Grav. \textbf{20}, R301 (2003)
%%%%%%%%%%%%%%%%%%%%%%%%%%
\bibitem{app-astro-3}
J. L. Wright, PhD thesis, University of Glasgow (2022)
%%%%%%%%%%%%%%%%%%%%%%%%%%%
\bibitem{Ayala:2021osy}
A.~Ayala, L.~A.~Hern\'andez, K.~Raya and R.~Zamora,
Phys. Rev. D \textbf{103}, 7, 076021 (2021)
[erratum: Phys. Rev. D \textbf{104}, 3, 039901 (2021)]
%%%%%%%%%%%%%%%%%%%%%%%%%%%%%%%%%%%%
\bibitem{Kapusta:2006pm}
J.~I.~Kapusta and C.~Gale,
\textit{Finite-temperature Field Theory: Principles and Applications},
Cambridge University Press, 2011
%%%%%%%%%%%%%%%%%%%%%%%%%%%%%%%%%%%%
\bibitem{Laine:2016hma}
M.~Laine and A.~Vuorinen,
\textit{Basics of Thermal Field Theory},
Lect. Notes Phys. \textbf{925},
Springer, 2016
%%%%%%%%%%%%%%%%%%%%%%%%%%%%%%%%%%%%
\bibitem{Abramowitz2006}
M.~Abramowitz and I.~A.~Stegun,
\textit{Handbook of Mathematical Functions},
US Dept. of Commerce, 1972
%%%%%%%%%%%%%%%%%%%%%%%%%%%%%%%%%%%%

%%%%%%%%%%%%%%%%%%%%%%%%%%%%%%%%%%%%

%%%%%%%%%%%%%%%%%%%%%%%%%%%%%%%%%%%%

%%%%%%%%%%%%%%%%%%%%%%%%%%%%%%%%%%%%

%%%%%%%%%%%%%%%%%%%%%%%%%%%%%%%%%%%%
\end{thebibliography}
\end{document}